\documentclass[apjl]{emulateapj}

\usepackage{natbib}
\usepackage{xcolor}
\usepackage[english]{babel}
\usepackage{graphicx}

\slugcomment{accepted for publication in ApJ Letters}
\shorttitle{Reproducing Andromeda's Thin Plane of Satellites}
\shortauthors{Buck, Macci\`o, Dutton}

\begin{document}

\title{Evidence for Early Filamentary Accretion from the Andromeda Galaxy's Thin
Plane of Satellites}

\author{Tobias Buck\altaffilmark{1}, Andrea V. Macci\`o\altaffilmark{2} and Aaron A. Dutton\altaffilmark{3}}
\affil{Max-Planck Institut f\"ur Astronomie, K\"onigstuhl 17, D-69117 Heidelberg, Germany}

\altaffiltext{1}{buck@mpia.de}
\altaffiltext{2}{maccio@mpia.de}
\altaffiltext{3}{dutton@mpia.de}

\begin{abstract}
Recently it has been shown that a large fraction of the dwarf
satellite galaxies orbiting the Andromeda galaxy are surprisingly
aligned in a thin, extended and kinematically coherent planar
structure. The presence of such a structure seems to challenge the
current Cold Dark Matter paradigm of structure formation, which
predicts a more uniform distribution of satellites around central
objects.  We show that it is possible to obtain a thin, extended,
rotating plane of satellites resembling the one in Andromeda in
cosmological collisionless simulations based on the Cold Dark Matter
model.  Our new high resolution simulations show a correlation between
the formation time of the dark matter halo and the thickness of the
plane of satellites. Our simulations have a high incidence of
satellite planes as thin, extended, and as rich as the one in
Andromeda and with a very coherent kinematic structure when we select
high concentration/early forming halos. By tracking the formation of
the satellites in the plane we show that they have been mainly
accreted onto the main object along thin dark matter filaments at high
redshift. Our results show that the presence of a thin, extended,
rotating plane of satellites is not a challenge for the Cold Dark
Matter paradigm, but actually supports one of the predictions of this
paradigm related to the presence of filaments of dark matter around
galaxies at high redshift.
\end{abstract}


\section{Introduction}

The success of the currently favored model of structure formation, the
Cold Dark Matter model, lies in its outstanding accordance with
observations on large scales \citep{T04,S05}. However, over the last
years several problems on galactic and sub galactic scales have been
reported, e.g.  the missing satellite \citep{K99,M99} and cusp-core
\citep{M94,B11} problems. A potential solution to these issues lies in
an inappropriate comparison between observations and simulations which
consider only the dark matter component. Cosmological simulations
including baryonic physics (gas and stars) have been shown to be able
to ease those problems and to bring the CDM model in agreement with
observations \citep{B07,Maccio10,G10,D14,b14}.

Recent observations of dwarf satellite galaxies around Andromeda
\citep{MC06,KG06,I13nature}, and possibly also around the Milky Way
\citep{M08}, have suggested these satellites to align in a
rotationally supported disc. \citet{I13nature} find that 15 out of 27
satellites in the Pan-Andromeda Archaeological Survey
\citep[PAndAS,][]{PANDAS} lie in a thin plane of thickness ($12.6 \pm
0.6$) kpc. Furthermore, using line-of-sight velocities, they report 13
of the satellites in the  plane to corotate.  This kind of spatial and
kinematic alignment is not easily found in Cold Dark Matter
simulations \citep{P14,P14_2,I14}, though not impossible
\citep{L09,I14}. Furthermore there are other studies  which have
searched for planes of satellites in hydrodynamical simulations
\citep[][and references therein]{Lovell11,G14}.  There have been
indications that filamentary accretion of subhalos can  lead to
anisotropic spatial distributions of them \citep{L09,Lovell11,L14}.
Nevertheless none of the previous studies was able to closely
reproduce the parameters of Andromeda's plane of satellites  (number
of members, thickness, corotation, and extension).  Since the
accretion and distribution of dwarf galaxies is primarily governed by
the global gravitational potential, which is dominated by the dark
matter distribution, this issue has raised (again) the question
whether the Cold Dark Matter model is correct or, conversely, needs to
be revised.

In this letter we present new high resolution ``zoom-in'' dark matter
only simulations of Andromeda sized dark matter halos which reveal
thin rotating planes of sub halos like in the case of Andromeda.  The
key insight that enables us to solve this puzzle is that halos formed
through filamentary accretion should form earlier and thus can be
selected by their higher than average present day  dark halo
concentrations.

This paper is organized as follows: In \S2 we present the simulations,
including the host halo selection criteria and our plane finding
algorithm.  In section \S3 we then present the results of our study
and in \S4 we present our conclusions.

\begin{figure*}
\includegraphics[width=.49\textwidth]{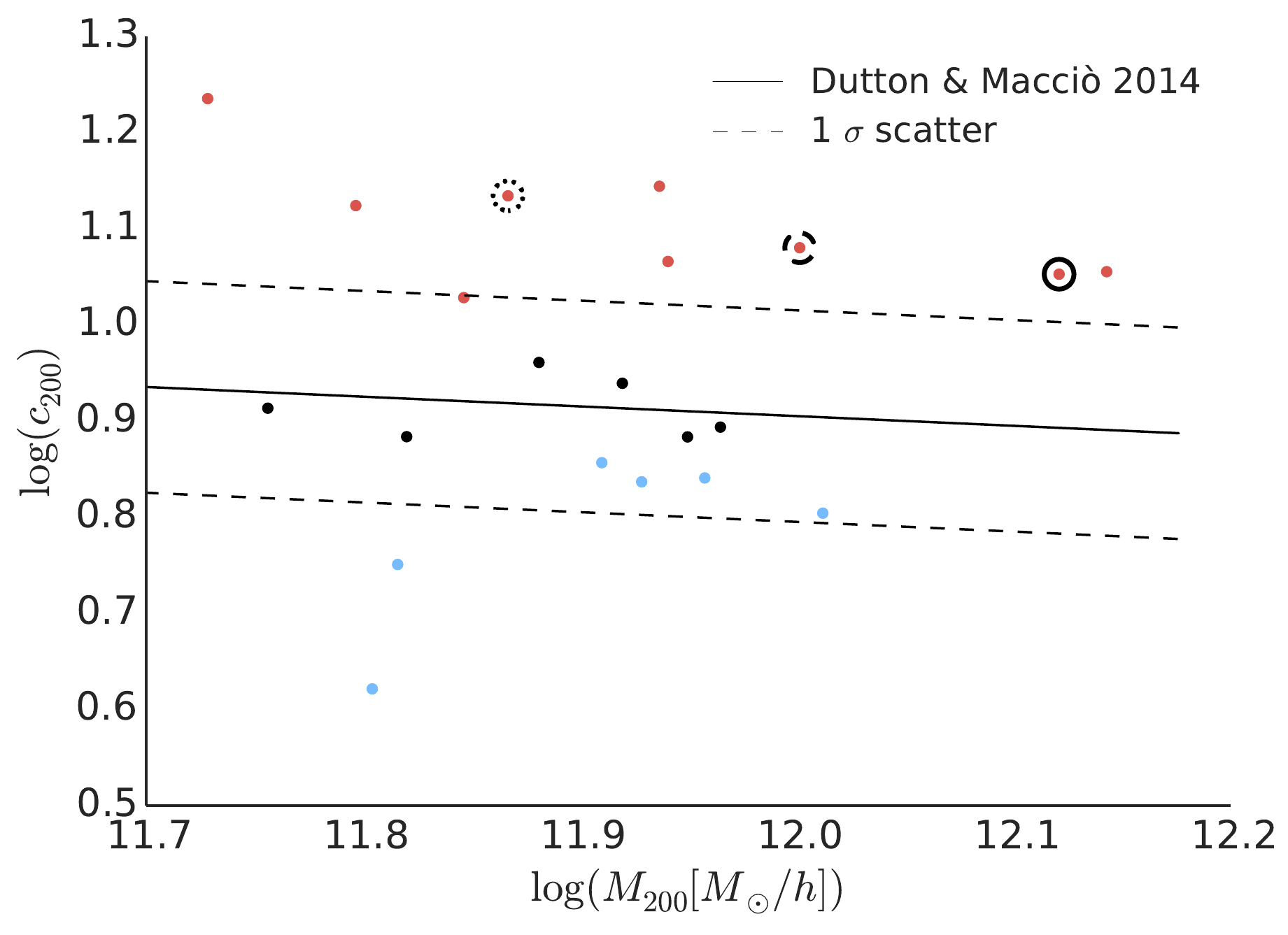}
\includegraphics[width=.49\textwidth]{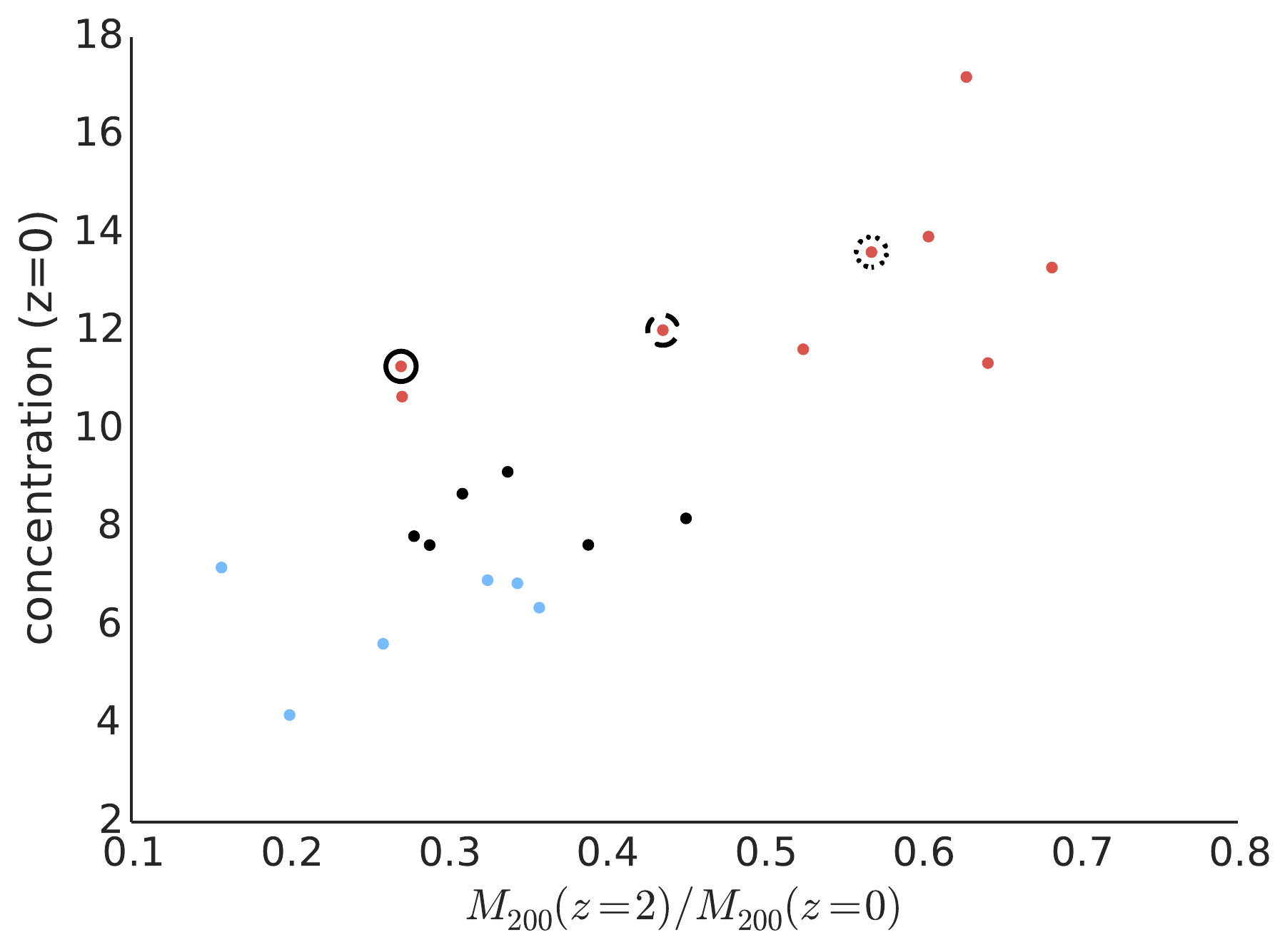}
\caption{\emph{Left panel}: Concentration mass relation. This diagram
  shows the concentration as a function of mass of the high resolution
  halos. The solid line is the average relation from
  \citet{Dutton}. The dashed line indicates the $1\sigma$ scatter of
  this relation. Color coding shows the division of the halos into
  high, average and low concentration. \emph{Right panel}: Mass growth
  vs. concentration. The plot shows the concentration at $z=0$ as a
  function of mass at $z=2$ in terms of the present day mass. Strongly
  growing halos (late forming) are located on the left while least
  growing halos (early forming) are located on the right. The color
  coding shows the division of the halos into high, average and low
  concentration. The halos with satellite planes coming closest to the
  values of Andromeda are marked by black circles. The solid one is
  \emph{halo A}, the dashed one is \emph{halo B}.}
\label{fig:conc}
\end{figure*}

\section{Simulations}

We performed 21 high resolution ``zoom-in'' dark matter only
simulations using the N-body code {\sc{pkdgrav2}} \citep{S,S13}. We
select Andromeda-like mass halos ($5\times10^{11} <
M_{200}/[h^{-1}M_{\odot}] < 1.5\times10^{12}$), where the halo mass is
defined with respect to 200 times the critical density of the
universe.  The halos were selected from three cosmological boxes of
sides 30, 60 and 80 $h^{-1}$Mpc from \citet{Dutton} which used
cosmological parameters from the Planck Collaboration
\citeyearpar{Planck}: $\Omega_m=0.3175$, $h=0.671$, $\sigma_8=0.8344$,
$n=0.9624$. 

Initial conditions for zoom-in simulations were created using
a modified version of the grafic2 package \citep{B01} as described in
\citep{Pz14}. The refinement level was chosen to maintain a roughly
constant relative resolution, e.g. $\sim 10^7$ dark matter particles
per halo with particle masses of $\sim 10^5 h^{-1}$M$_{\odot}$. This
allows us to reliably resolve substructure down to $\sim 5\times 10^7
h^{-1}$M$_{\odot}$.  The force softening of our simulations ranges
between $0.25$ $h^{-1}$ kpc and $0.36$ $h^{-1}$ kpc.

\subsection{Halo concentration and formation time}
Aside from halo mass, the only other selection criteria for our
objects is the concentration, which is a proxy for halo formation time
\citep{W02}. The right panel of Fig.~\ref{fig:conc} shows the
concentration at $z=0$ as a function of the mass growth since $z=2$.
The clear correlation validates our approach of using the
concentration as a first proxy for  the halo formation time.

The reasoning behind such a choice is that, at a {\bf{fixed mass}} at the
present time, early forming halos are more likely to form at the
nodes of intersection of a few filaments of the cosmic web, while
typical halos tend to reside inside such filaments \citep{D09}. One
then might expect that, rare, early forming halos would accrete
satellites from a few streams that are narrow compared to the halo
size, while typical halos accrete satellites from a wide angle in a
practically spherical pattern.

In the left panel of Fig.~\ref{fig:conc} we show the
concentration-mass relation of our high resolution halos.  Here the
concentration is defined as $c_{200}=R_{200}/r_{-2}$, where $R_{200}$
is the virial radius, and $r_{-2}$ is the radius where the logarithmic
slope of the density profile is $-2$.  We select roughly equal number
of high (red points), average (black points) and low (blue points)
concentration halos (see Fig.~\ref{fig:conc}).

The solid line is the power law fit from \cite{Dutton}, while the
dashed ones show the $1\sigma$ intrinsic scatter of 0.11 dex around
the mean. Our high concentration halos have on average an offset of
about $2\sigma$ from the mean relation. This means these halos are
the rarest 2.3\% of the whole population.  In a random sample of
halos it would thus require $\sim 40$ simulations to recover such
rare halo. This helps to explain why previous high resolution
simulations were unable to reproduce the observed properties of the
satellite distribution around the Andromeda galaxy: they simply did
not sample enough halos to find the rarer earliest forming ones.

\subsection{Satellite selection}
Our simulations reveal hundreds of resolved subhalos which have to be
matched to actual luminous galaxies. Galaxy formation models robustly
predict the luminous subhalos to be the ones most massive at infall
times \citep{K04,C06,V06}. Thus, we select a sample of the 30 most
massive sub halos at the time of the accretion, where we restrict our
analysis to sub halos within the virial radius of the host halo
($\sim 200$ kpc).

Although observations around Andromeda use a special selection
function given by the peculiarities of the Pan-Andromeda
Archaeological Survey (PAndAS) \citep{PANDAS}, we choose not to reproduce the selection
function for a number of reasons. Firstly, it requires
surface-brightness information which we do not have in our (dark
matter only) simulations. Secondly, the PAndAS footprint is unique to
the Andromeda galaxy, being non-circular, and including a region
around its most massive satellite M33. Thus it would not make sense to
apply the same footprint to a cosmological simulation. Thirdly, the
spatial depth of the survey is somewhat uncertain due to the
difficulty of measuring accurate distances to the satellites. Rather,
we apply a simple, reproduceable, and physically motivated selection
criteria.  As our satellite population we select the most massive sub
halos (at the time of infall) within the $z=0$ virial radius, $R_{\rm
  vir}$. Choosing satellites within the virial radius leaves us with a
bigger volume ($\approx R^3_{\rm vir }$) compared to the observations
and hence we use a number of 30 satellites instead of 27. Furthermore,
there is some arbitrariness in the number of satellites related to
Andromeda. Nine known satellites (two that lie inside the PAndAS
field, and seven that lie outside) were not considered by
\citet{I14Apj,Conn}.  Nevertheless we experimented with sample sizes of
25, 27 and 30 satellites and found no major differences in the plane
statistics. 

\subsection{Plane finding algorithm}
In order to find planes in the distribution of sub halos we generate
a random sample of planes defined by their normal vector. All planes
include the center of the main halo. To uniformly cover the whole
volume we generate 100,000 random planes with a fixed thickness of
2$\Delta=40h^{-1}$ kpc. After specifying a plane we calculate the
distance of every satellite to this plane. A satellite is considered
to lie in the plane if its distance to the plane is smaller than
$\Delta$. For each plane we calculate the number of satellites in the
plane and its root-mean-square thickness $\Delta_{\rm rms}$. We then
select for every number of satellites in the plane the one which is
thinnest and richest to analyze for kinematics \citep[for further details
see also][]{G14,I14Apj,Conn}. 

The plane of satellites around Andromeda can be characterized by 4
parameters including the number of satellites in the plane ($N_{\rm
  in}$), the number of corotating satellites ($N_{\rm corot}$), the
thickness of the plane ($\Delta_{\rm rms}$) and its extension.
For Andromeda these values are $N_{\rm in}=15$, $N_{\rm corot}=13$,
$\Delta_{\rm rms}= 12.6 \pm 0.6$ kpc, and a projected diameter of
$\sim 280$ kpc.

\section{Results}
\subsection{Plane thickness vs halo concentration}

\begin{figure}
\includegraphics[width=1.0\columnwidth]{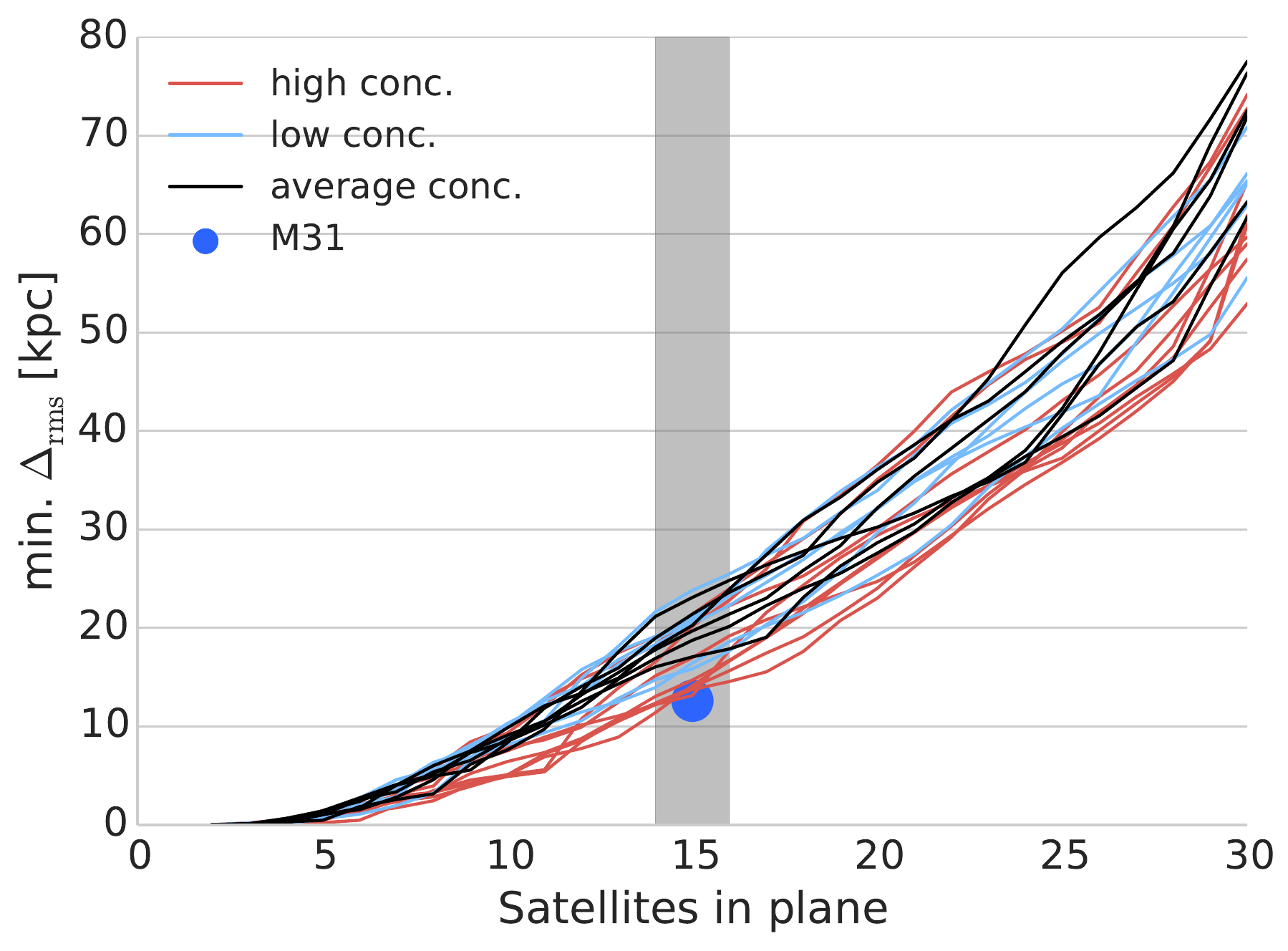}
\caption{Minimal root-mean-square thickness, $min.\ \Delta_{\rm rms}$,
  of planes as a function of number of satellites in the plane. Each
  line represents a different dark matter halo. \emph{Red} lines show
  high concentration halos, \emph{blue} lines show low concentration
  halos and \emph{black} lines show average concentration halos. The
  thinnest planes occur in the highest concentration halos.  The
  \emph{blue dot} shows the rms value of the plane of satellites
  observed around Andromeda (\cite{I13nature,Conn}). The \emph{grey area}
  represents a nominal uncertainty in the number of satellites in the plane
  around Andromeda of $\pm$1.}
\label{fig:rms}
\end{figure}

In a first step of our analysis we investigated the correlation
between concentration (as a proxy of halo formation time) and
thickness of the plane. Fig.~\ref{fig:rms} shows the plane thickness
as a function of the number of satellites in the plane, with lines
color coded according to halo concentration. There is a clear
dependence of plane thickness on the concentration of the halo. The
thinnest planes are only found in the highest concentration (red
lines), and hence earliest forming halos. Furthermore, only high
concentration halos have planes as thin as observed in Andromeda
(assuming 15 members).  The smooth relation between plane thickness
and number of satellites in the plane suggests that there is an
arbitrariness in the number of satellites chosen to be in the
plane. We do not see clear evidence of two distinct spatial structures
such as a planar and spherical distribution of satellites. We further
note that an investigation of the satellite distribution does not
reveal a more concentrated satellite distribution in high
concentration halos, which might trivially explain the dependence of
plane thickness on concentration that we find.

\subsection{Corotation vs number of satellites in plane}

\begin{figure*}
\begin{center}
\includegraphics[width=\textwidth]{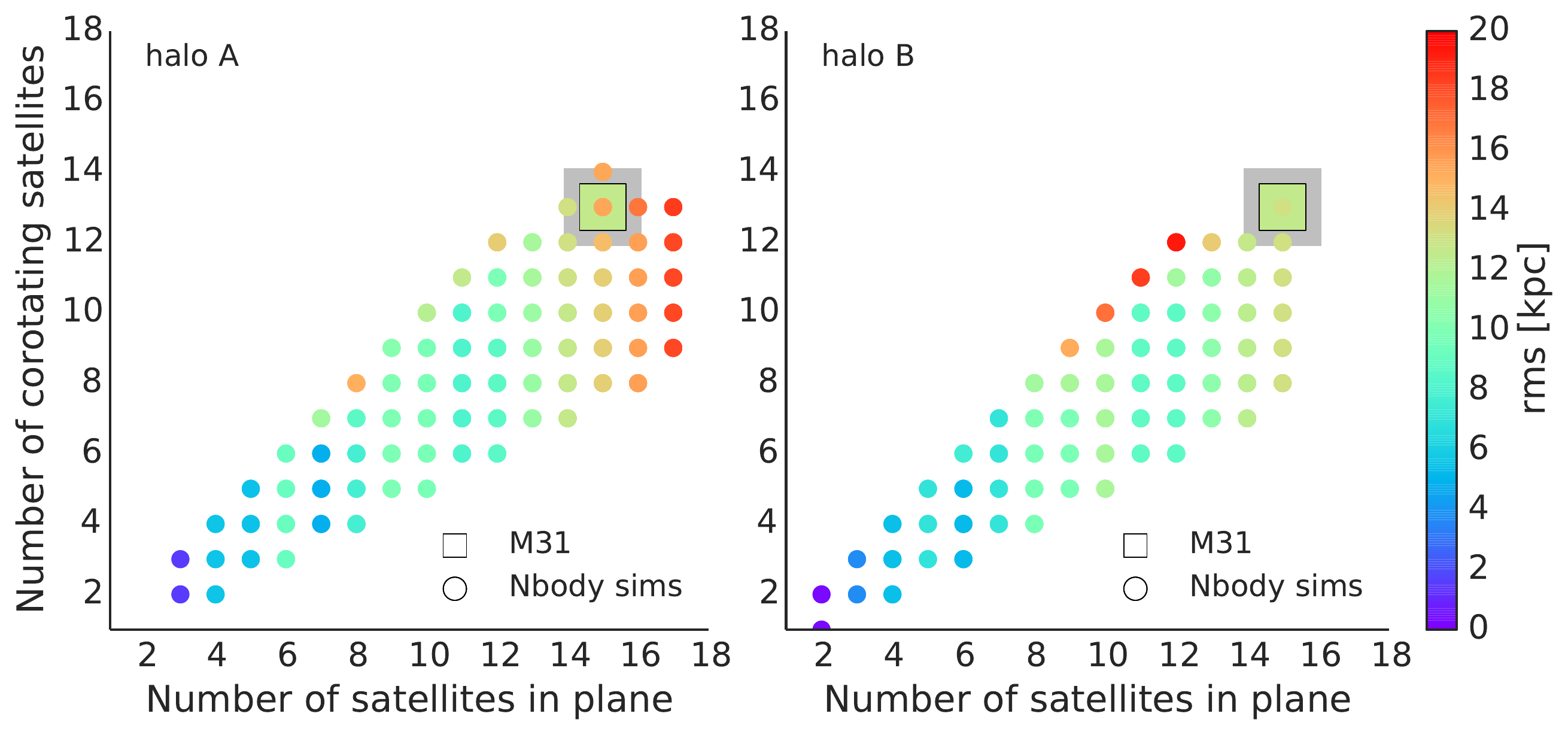}
\end{center}
\caption{Number of corotating satellites vs. number of satellites in
  the plane for the two best matching high concentration halos ({\it
    halo A (left) \& halo B (right)}). The points are color coded by
  the rms thickness ($\Delta_{\rm rms}$)  of each plane. The
  \emph{square} marks the values observed for Andromeda (15 in plane,
  13 corotating), where the \emph{grey} shaded area marks the
  uncertainty of $\pm$ 1 satellite in the plane and $\pm$ 1 corotating
  satellite. The \emph{dots} show the values of the planes found for
  this halo. }
\label{fig:corot}
\end{figure*}

Fig.~\ref{fig:corot} shows the outcome of our plane finding algorithm
for the two (early forming) halos best matching the values of
Andromeda (halo A: $N_{\rm in}=15, N_{\rm corot}=14$, $\Delta_{\rm
  rms}=15.8$ kpc, diameter $\sim 220$ kpc; and halo B: $N_{\rm in}=15,
N_{\rm corot}=13$, $\Delta_{\rm rms}=12.9$ kpc, $\sim 450$ kpc).  The
plots show for every value {\bf{of the number}} of satellites in the plane,
$N_{\rm in}$, the number of corotating satellites, $N_{\rm
  corot}$. Every dot in the plot represents a different plane.  In
both halos one can find planes with up to 12 members that have a 100\%
corotation fraction. One can always find planes with no coherent
kinematics (i.e., a corotation fraction of 50\%).  The points are
color coded according to the thickness of the plane $\Delta_{\rm
  rms}$.  As would be expected, planes with more satellites tend to be
thicker.  The thickness of the plane is also to first order
independent of the corotation fraction, except at the highest
corotation fractions.

\begin{figure*}
\centering
\includegraphics[width=.8\textwidth]{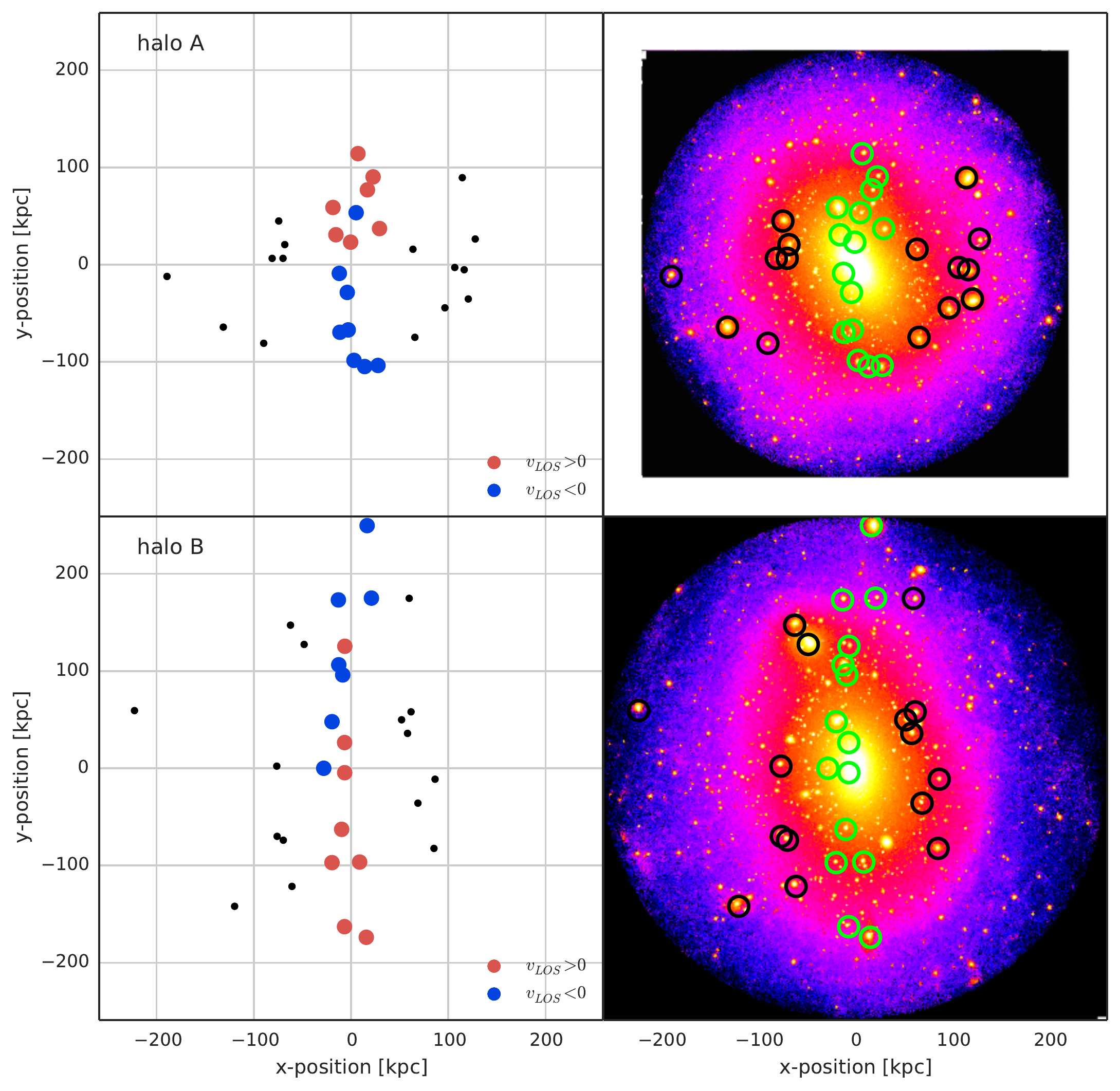}
\caption{Edge on view of planes. The \emph{left} panels show the sign
  of the line-of-sight velocity of the best matching plane consisting
  of 15 satellites (colored dots).  Black dots show the satellites not
  in the plane. The \emph{right} panels show high resolution density
  plots of the halo hosting the best matching plane. \emph{Green
    circles} indicate the sub halos in the plane and \emph{black
    circles} indicate sub halos not in the plane. \emph{The upper
    panels} show \emph{halo A}, while the \emph{lower panels} show
  \emph{halo B} a slightly more massive halo, also revealing the
  dependence of radial extension on mass.}
\label{fig:images}
\end{figure*}

This figure also shows that there is some arbitrariness in choosing
the {\it best} plane from a simulation. For example for halo A (left
panel in Fig.~\ref{fig:corot}, and upper panel of
Fig.~\ref{fig:images}), we can find a plane with 15 satellites and 14
corotating (one more than Andromeda).  If we restrict to a plane of 14
satellites (instead of 15) we still find a large corotating fraction
($N_{\rm corot} = 13$) but in a thinner plane $\Delta_{\rm rms} =
13.9$ kpc, which is consistent with the Andromeda value.

We are also able to find a thin plane even richer in number of
satellites and coherent motion as the one found around Andromeda.  For
example, in {\it halo A} we are able to find planes consisting of up
to 17 satellites from which 13 share the same rotation direction (see
left panel in Fig.~\ref{fig:corot}). 

\subsection{Visual impression of planes matching Andromeda}

Fig.~\ref{fig:images} shows one particular projection of two early
forming halos with plane parameters coming closest to the ones
observed around Andromeda.  The left hand side shows the spatial
extension and the kinematics of the system, clearly revealing a thin
but radially extended plane with coherent motion of the satellites
around their host. The right hand side shows a dark matter density map
of the host halos, with superimposed the two satellite populations,
one in the plane (green circles) and one outside the plane (black
circles). Such distinct planes were reported before in CDM simulations
\citep{G14,I14,P14} but never as rich as the ones found here. 

It is worth noticing that previous simulation studies either were
limited by the capability of resolving enough (sub)structures, e.g.,
the Millennium II simulation \citep{Sp09} or, conversely, were limited
by very low number statistics of host halos \citep{G14}, moreover the
halos were not selected according to formation time.  Since we are
able to find planes as rich as the one around Andromeda in at least 3
out of our 20 simulations, the {\bf{rarity}} of the planes can be explained
by the rareness of early forming halos.

\subsection{Formation scenario}

\begin{figure*}
\centering
\includegraphics[width=.8\textwidth]{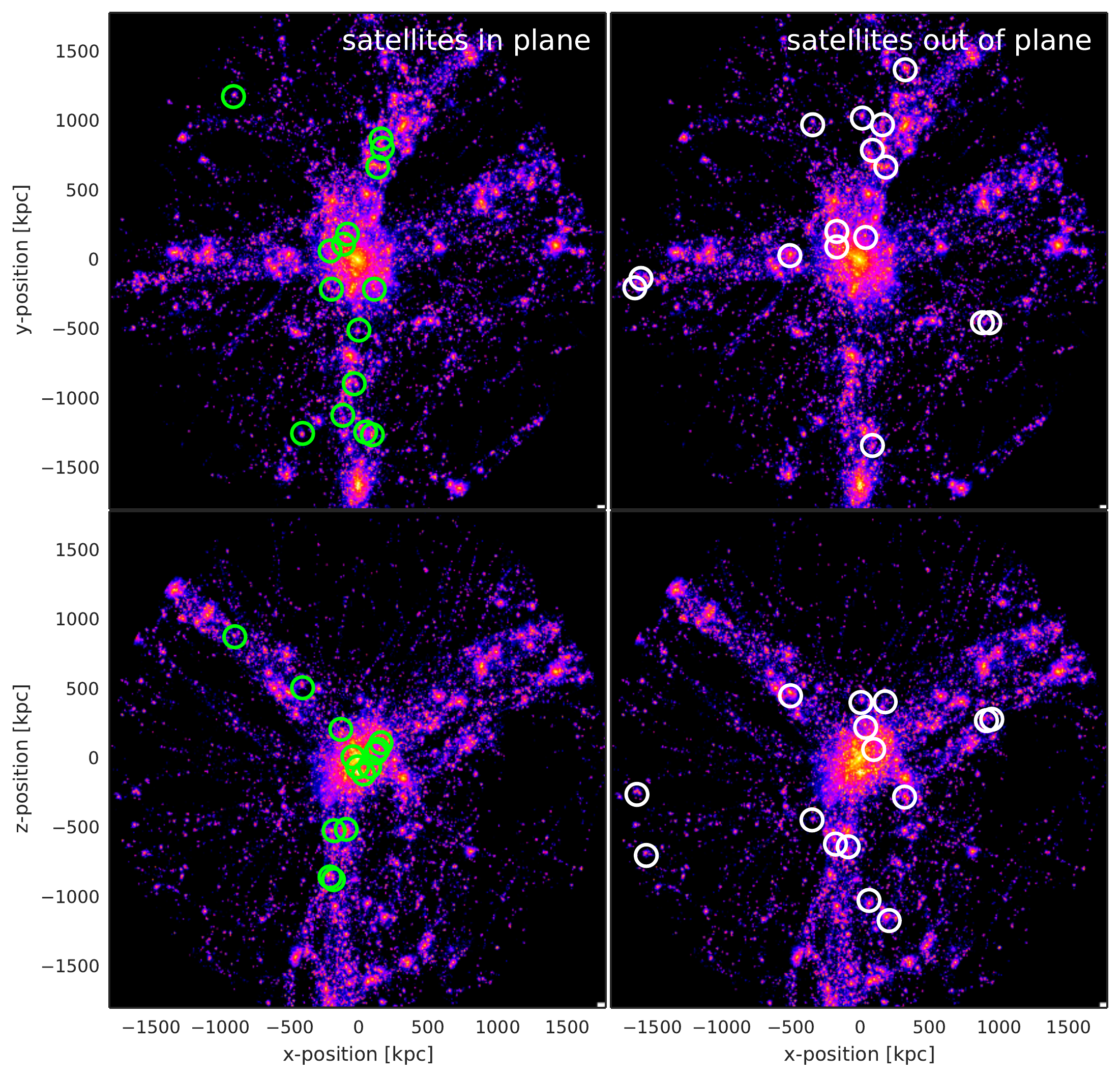}
\caption{High redshift ($z=3$) density plots of satellite distribution
  ending up in the plane and outside of the plane of \emph{halo
    A}. The \emph{left panel} shows density plots of $x-y$ and
  $x-z$-projections of the satellites ending up in the plane at $z=0$
  while the \emph{right panel} shows density plots of the $x-y$ and
  $x-z$-projections of the satellites ending up outside of the
  plane. The upper panel shows that satellites ending up in the plane
  are accreted along two filaments coming from opposite sides of the
  main halo, which set a preferred infall direction. While satellites
  not ending up in the plane are accreted from everywhere. Comparison
  with the lower panel shows clearly that satellites ending up in the
  plane lie within the filaments such that their projection collimates
  in the center of the halo while halos not ending up in the plane
  scatter around the main halo indicating that they are not part of
  the filaments.}
\label{fig:filaments}
\end{figure*}

We now move to the question where does this spatial and kinematic
coherence come from. By tracing the satellites in the plane back to a
redshift of $z=3$ we reveal the accretion of the satellites are along
dark matter filaments. This can be seen in Fig.~\ref{fig:filaments}
where we show a density plot of the main halo and its substructure at
redshift $z=3$ indicating the satellites in the plane by green circles
and the satellites outside the plane by white circles. Providing two
different projections these plots prove the connection between
accretion along filaments and the property of being in a kinematically
coherent plane at redshift $z=0$.

The presence of a plane of satellites in Andromeda seems then to
suggest an (unusual) early formation epoch for this galaxy. Such a
scenario is consistent with other observational evidences. For its
stellar mass ($\sim 10^{11}$M$_\odot$) Andromeda lives in a lower mass
dark matter halo than typical galaxies of the same mass \citep{M10}. At
these stellar masses the majority of galaxies are bulge-dominated and
non star forming, while Andromeda is disk-dominated and star forming,
consistent with an early mass accretion history, devoid of recent
major mergers. A similar line of reasoning also suggests an early
formation epoch for the Milky Way.

\section{Conclusion}

We have explored the connection between the formation time of a host
dark matter halo and the alignment and coherent kinematics  of its
subhalos using 21 high-resolution (10 million particles) cosmological
N-body simulations. Our key new result is that high concentration
(earlier forming) halos tend to have thinner and richer planes.  Our
simulations show that the presence of a thin, rotating, and extended
plane of satellites like the one observed around the Andromeda galaxy
is not a challenge for the Cold Dark Matter paradigm. Conversely it
supports one of the key predictions of such a model, namely the
presence of large filaments of dark matter around galaxies at high
redshift and the web-like nature of cosmic structures in the Universe.

The connection we report in this letter between the formation time with
the satellite distribution at the present time, opens a new
possibility of constraining the topology of the dark matter accretion
pattern.

\acknowledgments
\section*{Acknowledgments}
The authors acknowledge support from the
   Sonderforschungsbereich SFB 881 “The Milky Way System” (subproject
   A2) of the German Research Foundation (DFG).  Simulations have been
   performed on the THEO clusters of the Max-Planck-Institut fuer
   Astronomie at the Rechenzentrum in Garching.  We thank R. van den
   Bosch and F. Walter for their comments on an early version of this
   draft.
   
\newpage
\bibliography{bib}

\end{document}